\documentclass[aps,pre,twocolumn,showpacs,floatfix]{revtex4}

\usepackage{graphicx}
\usepackage{amsmath,amssymb}
\usepackage{mathtools}
\usepackage{xcolor}
\usepackage[enableskew]{youngtab}
\allowdisplaybreaks

\begin{document}
	
\title{Power series solution of the inhomogeneous exclusion process}
	
\author{Juraj Szavits-Nossan}
\email{jszavits@staffmail.ed.ac.uk}
\affiliation{SUPA, School of Physics and Astronomy, University of Edinburgh, Peter Guthrie Tait Road, Edinburgh EH9 3FD, United Kingdom}
	
\author{M. Carmen Romano}
\affiliation{SUPA, Institute for Complex Systems and Mathematical Biology, Department of Physics, Aberdeen AB24 3UE, United Kingdom}
\affiliation{Institute of Medical Sciences, University of Aberdeen, Foresterhill, Aberdeen AB24 3FX, United Kingdom}

\author{Luca Ciandrini}
\affiliation{DIMNP, Universit\'{e} de Montpellier, CNRS, Montpellier, France}
\affiliation{L2C, Universit\'{e} de Montpellier, CNRS, Montpellier, France}

\date{\today}
	
\begin{abstract}
We develop a power series method for the nonequilibrium steady state of the inhomogeneous one-dimensional totally asymmetric simple exclusion process (TASEP) in contact with two particle reservoirs and with site-dependent hopping rates in the bulk. The power series is performed in the entrance or exit rates governing particle exchange with the reservoirs, and the corresponding particle current is computed analytically up to the cubic term in the entry or exit rate, respectively. We also show how to compute higher-order terms using combinatorial objects known as Young tableaux. Our results address the long outstanding problem of finding the exact nonequilibrium steady state of the inhomogeneous TASEP. The findings are particularly relevant to the modelling of mRNA translation in which the rate of translation initiation, corresponding to the entrance rate in the TASEP, is typically small.
\end{abstract}
	
	

	
	
	
\maketitle

\section{Introduction}
\label{introduction}

The exclusion process is a stochastic driven lattice gas of particles interacting by excluded volume. It appears as a model for diverse transport phenomena including mRNA translation~\cite{MacDonald68}, enzyme kinetics~\cite{Shapiro82}, molecular motors~\cite{Lipowsky01,Parmeggiani03} and vehicle traffic~\cite{NagelSchreckenberg92}. Here we are interested in an exclusion process in which particles move unidirectionally along a discrete lattice in contact with two particle reservoirs. This model is called the totally asymmetric simple exclusion process (TASEP), whose distinctive feature is a nonequilibrium steady state that carries a net current of particles.

A fundamental difference between equilibrium and nonequilibrium steady states is that the former are independent of the dynamics (provided the detailed balance is respected) while the latter are not. As a result, nonequilibrium steady states are in general unknown, even in one-dimensional systems~\cite{Privman97}. An important exception is the TASEP in which particles move along the lattice at constant rate, for which the steady state is known explicitly~\cite{Shapiro82,Derrida92,DEHP}. The exact solution reveals intriguing boundary-induced phase transitions~\cite{Krug91} that have no equilibrium counterpart.

However, the steady state of the \emph{inhomogeneous} TASEP, in which particles move at position-dependent rates (also called disordered TASEP or TASEP with sitewise disorder),  is a long outstanding unsolved problem in nonequilibrium statistical physics. The majority of work on the inhomogeneous TASEP concerns the phenomenon of phase separation induced by isolated inhomogeneities~\cite{Janowsky92,Schutz94,Janowsky94,Kolomeisky98,Schmidt15,Scoppola15} and full disorder~\cite{Tripathy97,Krug00,Harris04,Shaw04,Greulich08}, which is typically studied using numerical simulations and various types of mean-field approximations that neglect correlations between neighbouring sites~\cite{MacDonald68,Janowsky92,Kolomeisky98,ChouLakatos04,Shaw04}. Importantly, there is a long-standing interest in the inhomogeneous TASEP as a model for mRNA translation, in which ribosomes progress along the mRNA at codon-dependent hopping rates~\cite{MacDonald68,Heinrich80,Shaw04,Mitarai08,Romano09,Zia11,Reuveni11,Ciandrini13,Tuller16}.

Motivated by this lack of exact results, one of us recently developed a power series method for the steady state of the inhomogeneous TASEP with binary disorder in which site-dependent hopping rates are either $r<1$ (``slow'' sites)  or $1$ (``normal'' sites)~\cite{SzavitsNossan13}. The power series was performed up to the quadratic order in the variable $r$ for lattices with one and two slow sites. More recently, we extended this method to a TASEP-like model of mRNA translation that includes codon-dependent elongation rates, two-step hopping mechanism and extended particles that cover $10$ lattice sites~\cite{SNCR18}. The power series was preformed up to the quadratic order in the rate of translation initiation $\alpha$, which is typically rate-limiting for translation under physiological conditions~\cite{Ciandrini13, Kudla09,Salis09} and therefore, it crucially reduces contact between particles~\cite{Duc18}.

In this paper we develop a power series method for the steady state of the inhomogeneous TASEP with open boundary conditions and site-dependent hopping rates in the bulk. We perform a power series in the entrance and exit rates $\alpha$ and $\beta$ and find an expression for the current up to the cubic order. In addition, we show how to compute higher-order terms using combinatorial objects called Young tableaux of shifted shape~\cite{Adin15}. Interestingly, the connection between the inhomogeneous TASEP and Young tableaux that we establish here is different from the one in Ref.~\cite{Corteel07}, which applies only to the homogeneous TASEP. Our method is robust and applicable to many other TASEP-like models, especially ones that describe mRNA translation in which the entrance rate is typically small.

The paper is organised as it follows. The model is described in Section \ref{model} and its exact steady state is formally derived in Section \ref{exact_solution}. The main idea behind the power series method is summarised in Section \ref{main_idea}. Sections \ref{small_alpha} and \ref{small_beta} describe the power series in $\alpha$ and $\beta$, respectively. Main results are summarised in Section \ref{conclusion}.

\section{Inhomogeneous exclusion process}
\label{exclusion_process}

\subsection{The model}
\label{model}

\begin{figure}[hbt]
\centering\includegraphics[width=7cm]{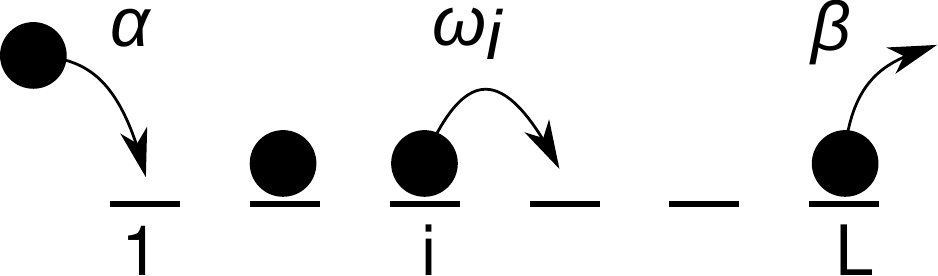}
\caption{Schematic of the TASEP with entrance rate $\alpha$, site-dependent hopping rates $\omega_i$ for $i=1,\dots,L-1$ and exit rate $\beta$.}
\label{fig1}
\end{figure}

We consider the totally asymmetric simple exclusion process on a one-dimensional lattice consisting of $L$ sites, which is presented in Fig.~\ref{fig1}. The lattice is in contact with two particle reservoirs that allow particles to enter at site $1$ at rate $\alpha$ (provided the site $1$ is empty) and exit from site $L$ at rate $\beta$. Particles move unidirectionally along the lattice at site dependent rates $\omega_i$, provided that the site $i+1$ in front is not occupied by another particle.

Each lattice site $i$ is assigned an occupancy variable $\tau_i=0$ if the site is empty, and $\tau_i=1$ if it is occupied by a particle. Throughout the text we consider continuous-time dynamics with entrance rate $\alpha$, site-dependent hopping rates $\omega_i$, $i=1,\dots,L-1$, and exit rate $\beta$. The allowed transitions between configurations and their corresponding rates can be summarised as

\begin{subequations}
\begin{align}
& 0_1\overset{\alpha}{\rightarrow} 1_1\label{entrance}\\
& 1_i 0_{i+1}\overset{\omega_i}{\rightarrow} 0_i 1_{i+1},\quad i=1,\dots,L-1\label{hopping}\\
& 1_L\overset{\beta}{\rightarrow} 0_L,\label{exit}
\end{align}
\end{subequations}
where with $0_i$ ($1_i$) we represent the empty (occupied) site $i$.

The steady state of this process is determined by the following stationary master equation
\begin{align}
&\alpha(1-2\tau_1)P(0_1\tau_2\dots)-\beta(1-2\tau_L)P(\dots\tau_{L-1}1_L)+\nonumber\\
&\quad-\sum_{i=1}^{L-1} \omega_i(\tau_i-\tau_{i+1})P(\dots\tau_{i-1}1_i0_{i+1}\tau_{i+2}\dots)=0,
\label{master}
\end{align}
where $P(C)$ is the probability to find the system in the configuration $C=\tau_1\dots\tau_L$. The aim of this paper is to find $P(C)$ of an inhomogeneous exclusion process, which we then use to calculate the average particle current $J$, the local density $\rho_i$ and the total density $\rho$, which are defined as
\begin{align}
& J=\alpha\sum_{C}\left[1-\tau_1(C)\right]P(C)\,,\label{current}\\
& \rho_i=\sum_{C}\tau_i(C)P(C),\quad \rho=\frac{1}{L}\sum_{i=1}^{L}\rho_i \;.\label{density}
\end{align}

\subsection{Exact solution}
\label{exact_solution}

In this Section we present an exact solution for any stationary ergodic master equation, which serves as the basis for our power series method presented in Section \ref{series_expansion}. This solution is often overlooked in literature, because it is rarely practical for the reasons that we expose below. In that context, our power series method shows how to approximate $P(C)$ in a certain limit and extract information on the system described.

To this end, let us label configurations $C_i$, where $i$ runs from $1$ to $\mathcal{N}=2^L$, which is the total number of configurations. This allows us to rewrite the master equation (\ref{master}) in a more compact form,
\begin{equation}
M\mathbf{P}=\mathbf{0},
\label{master_compact}
\end{equation}
where $\mathbf{P}$ is a column vector of $\mathcal{N}$ steady-state probabilities $P(C_i)$ and $M$ is a $\mathcal{N}\times \mathcal{N}$ matrix whose elements $M_{ij}$ are given by
\begin{equation}
M_{ij}=\begin{cases}
W(C_j\rightarrow C_i), & i\neq j\\
-\sum_{k\neq i}W(C_i\rightarrow C_k), & i=j.
\end{cases}
\end{equation}
Here $W(C_i\rightarrow C_j)$ is the transition rate from $C_i$ to $C_j$ that takes one of the values in Eqs.~(\ref{entrance})-(\ref{exit}). 

There are two general methods for solving Eq.~(\ref{master_compact}) (and any other stationary ergodic master equation for that matter), one that uses determinants and the other that uses graphs; the latter is known as the Schnakenberg's network theory~\cite{Schnakenberg76} in physics or the matrix-tree theorem~\cite{Chaiken78} in mathematics.

\subsubsection{Solution using determinants}

We recall that the sum of all elements in each column of $M$ is zero, which means that the rows of $M$ are linearly dependent and the determinant of $M$ is zero. The Laplace expansion for the determinant $\textrm{det}M$ gives
\begin{equation}
\label{laplace_expansion}
0=\textrm{det}M=\sum_{i=1}^{\mathcal{N}}M_{ij}A_{i,j},
\end{equation}
where $A_{i,j}=(-1)^{i+j}\textrm{det}M^{(i,j)}$ is called $i,j$ cofactor of $M$ and $M^{(i,j)}$ is a matrix derived from $M$ by removing the $i$-th row and $j$-th column. Equation~(\ref{laplace_expansion}) can be rewritten as
\begin{align}
0=\textrm{det}M &=\sum_{i\neq j}M_{ij}A_{i,j}+M_{jj}A_{j,j}\nonumber\\
&=\sum_{i\neq j}(A_{i,j}-A_{j,j})M_{ij},
\label{eq:detM}
\end{align}
which means that $A_{i,j}=A_{j,j}$. In the last passage of Eq.~(\ref{eq:detM}) we used the definition of $M_{jj}=-\sum_{k\neq j}M_{kj}$. Inserting this result back into Eq.~(\ref{laplace_expansion}) and comparing it to Eq.~(\ref{master_compact}) shows that $P(C_i)$ is given by
\begin{equation}
\label{general_solution}
P(C_i)=\frac{\textrm{det}M^{(i,i)}}{\sum_{j=1}^{\mathcal{N}}\textrm{det}M^{(j,j)}}.
\end{equation}

Alternatively, we can solve Eq.~(\ref{master_compact}) using the Cramer's rule. Since the matrix $M$ is singular ($\textrm{det}M=0$), we can replace one of the equations in Eq.~(\ref{master_compact}) by the condition that $\sum_{i=1}^{\mathcal{N}}P(C_i)=1$. This new system of equations has a non-singular matrix, which can now be solved using the Cramer's rule.

\subsubsection{Solution using the Schnakenberg's network theory}

This method was first developed by Gustav Kirchhoff for electric circuits and was later refined by many others including the seminal Schnakenberg's paper~\cite{Schnakenberg76}. The reason we mention this method is because it will prove useful for our later analysis. 

For a given Markov jump process, let us call $V$ the set of all configurations $C_i$, $i=1,\dots,\mathcal{N}$ and $E$ the set of all pairs of configurations $(C_i,C_j)$ for which $W(C_i\rightarrow C_j)\neq 0$. The Schnakenberg's network theory connects this stochastic process to the directed weighted graph $G=(V,E)$ consisting of vertices $V$ and directed edges $E$ weighted by the corresponding transition rates $W$.

\begin{figure}[hbt]
\centering\includegraphics[width=7cm]{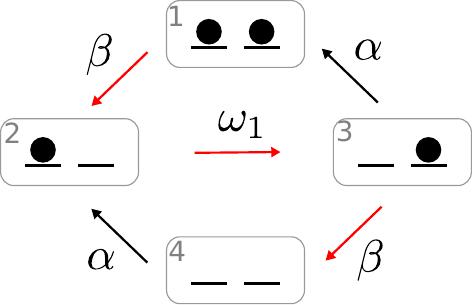}
\caption{Sketch of the graph connecting the configurations $C_i$ in a lattice of length $L=2$. The boxes represent all the possible configurations of the system, labelled by their index $i$ and connected by weighted edges. In red (colour online) we highlight the only spanning tree $T_4$ rooted at $C_4$ (bottom configuration). By following Eq.~(\ref{matrix-tree-theorem}), the probability of this configuration is proportional to $\beta^2\omega_1$.}
\label{fig2}
\end{figure}

Let us define a directed path in $G$ as a sequence of vertices $C_{s(1)},\dots,C_{s(n)}$ such that $(C_{s(i)},C_{s(i+1)})\in E$ for all $i=1,\dots,n-1$ \footnote{We used the index $s(i)$ for the $i$-th configuration in the sequence $S$ in order to remove any confusion with the indices $i=1,\dots,\mathcal{N}$ that we introduced earlier}. We define a spanning tree $T_i$ rooted at $C_i$ (also called spanning in-tree with a sink at $C_i$) as a subgraph of $G$ that contains all vertices in $V$ and only a subset of edges in $E$ such that there is exactly one directed path from every vertex $C_j\in V$ to $C_i$. An example of a spanning tree for a lattice of length $L=2$ is shown in Fig.~\ref{fig2}. Let us denote by $w(T_i)$ the weight of this spanning tree obtained by multiplying transition rates of all the edges in $T_i$. The following result, which is called the matrix-tree theorem, says that the minor $\textrm{det}(-M^{(i,i)})$ \footnote{The minus sign is here because of our definition of the matrix $M$, which is natural for stochastic processes; in the graph theory one usually works with $-M$ instead.} is equal to the sum of weights $w(T_i)$ of all spanning trees rooted at $C_i$

\begin{equation}
\textrm{det}(-M^{(i,i)})=(-1)^{\mathcal{N}-1}\textrm{det}M^{(i,i)}=\sum_{T_i}w(T_i)
\label{matrix-tree-theorem}
\end{equation}
Inserting Eq.~(\ref{matrix-tree-theorem}) into Eq.~(\ref{general_solution}) gives 
\begin{equation}
\label{Ptree}
P(C_i)=\frac{\sum_{T_i}w(T_i)}{\sum_{j=1}^{\mathcal{N}}\sum_{T_j}w(T_j)}.
\end{equation}

The main problem with Eqs.~(\ref{general_solution}) and (\ref{Ptree}) is that the number of terms increases exponentially with $L$, which makes finding $P(C_i)$ intractable. 

In the rest of this paper we develop a power series method that allows us to find $P(C_i)$ when either $\alpha$ or $\beta$ are small, without solving the full system in Eq.~(\ref{master_compact}).

\section{Power series of the steady state}
\label{series_expansion}

\subsection{Main method}
\label{main_idea}

The main method exploits the fact that $P(C)$ is a quotient of two multivariate polynomials. We can choose any rate, which we denote by $\kappa\in\{\alpha,\omega_1,\dots,\omega_{L-1},\beta\}$ and rewrite Eq.~(\ref{general_solution}) as
\begin{equation}
P(C)=\frac{f(C)}{\sum_{C'}f(C')}, \quad f(C)=\sum_{n=0}^{K(C)}f_n(C)\kappa^n.
\label{f}
\end{equation}
The unknown coefficients $f_n(C)$, $n=0,\dots,K(C)$, depend on the configuration $C$ and also on all the other rates, which we have omitted in order to simplify the notation. 

According to the Schnakenberg network theory, $\sum_{n=0}^{K(C)}f_n(C)\kappa^n$ is equal to the sum of weights $w(T_C)$ of all spanning trees $T_C$ rooted at $C$ that have exactly $n$ directed edges weighted by $\kappa$. The degree $K(C)$ is the maximum number of these edges over all spanning trees rooted at $C$. While in general we do not know the exact value of $K(C)$, we know that $K(C)$ is bounded from above by the maximum number of configurations (excluding $C$!) that have an outward edge weighted by $\kappa$. For example, if we choose $\kappa=\alpha$, then $K(C)\leq 2^{L-1}$ if $\tau_1(C)=1$ and $K(C)\leq 2^{L-1}-1$ if $\tau_1(C)=0$. 

In order to understand what Eq.~(\ref{f}) implies for the coefficients $f_n(C)$, we write the master equation (\ref{master}) in a generic form, 
\begin{equation}
\sum_{C'}W(C\rightarrow C')P(C)=\sum_{C'\neq C}W(C'\rightarrow C)P(C'),
\label{master_generic}
\end{equation}
where we denote by $e(C)$ the exit rate from $C$, i.e., 
\begin{equation}
e(C)=\sum_{C'}W(C\rightarrow C').
\end{equation}

For any two configurations $C$ and $C'$, we define $I_{C,C'}$ such that $I(C,C')=1$ if there is a transition from $C$ to $C'$ at rate $\kappa$ and $I_{C,C'}=0$ otherwise,
\begin{equation}
I_{C,C'}=\begin{cases}1 & W(C\rightarrow C')=\kappa\\
0 & \textrm{otherwise}.
\end{cases}
\end{equation}
We can now use $I_{C,C'}$ to decompose $W(C\rightarrow C')$ and $e(C)$ into  
\begin{align}
& W(C\rightarrow C')=W(C\rightarrow C')(1-I_{C,C'})+\kappa I_{C,C'},\label{W}\\
& e(C)=e_0(C)+\kappa\sum_{C'}I_{C,C'},\label{e}\\
& e_0(C)=\sum_{C'\neq C}W(C\rightarrow C')(1-I_{C,C'}).\label{e0}
\end{align}
Inserting Eqs.~(\ref{f}), (\ref{W}), (\ref{e}) and (\ref{e0}) into Eq.~(\ref{master_generic}) and collecting all the terms of order $\kappa^n$ gives the following recurrence relation for $f_n(C)$ in terms of $f_{n-1}(C')$
\begin{align}
& e_0(C)f_n(C)+\sum_{C'}I_{C,C'}f_{n-1}(C)=\sum_{C'}I_{C',C}f_{n-1}(C')\nonumber\\
&\quad +\sum_{C'}W(C'\rightarrow C)(1-I_{C',C})f_n(C').
\label{master_n}
\end{align}
The equation above is true for $n>0$. The terms containing $f_{n-1}(C)$ and $f_{n-1}(C')$ are absent for $n=0$, 
\begin{equation}
e_0(C)f_0(C)=\sum_{C'}W(C'\rightarrow C)(1-I_{C',C})f_0(C').
\label{master_0}
\end{equation}
By construction, the configurations whose all outward edges are weighted by $\kappa$ (so that $e_0(C)=0$) are completely absent from Eq.~(\ref{master_0}). This equation thus has a trivial solution $f_0(C)=0$, from which we conclude that $f_0(C)\neq 0$ only for configurations for which $e_0(C)=0$. This conclusion is crucial for our later analysis.

Once we solve Eq.~(\ref{master_n}) recursively up to some order $n$ we can Taylor expand $P(C)$ around $\kappa=0$
\begin{equation}
P(C)=\frac{\sum_{n=0}^{K(C)}f_n(C)\kappa^n}{\sum_{C'}\sum_{n=0}^{K(C')}f_n(C')\kappa^n}=\sum_{n=0}^{\infty}c_n(C)\kappa^n\,.
\label{Pseries}
\end{equation}
The coefficients $c_n(C)$ can be obtained by multiplying the denominator with the r.h.s. of Eq.~(\ref{Pseries}) and collecting all terms of the same order in $\kappa^n$. The coefficient $c_0(C)$ is simply given by
\begin{equation}
c_0(C)=\frac{f_0(C)}{\sum_{C}f_0(C)}.
\label{c0}
\end{equation}
Next, we define $b_n$ and $K_0$ as
\begin{align}
b_n &=\sum_{C}f_n(C), \quad n=0,\dots,K_0,\label{bn}\\
K_0 &=\underset{C}{\textrm{max}}\{K(C)\},\label{K0}
\end{align}
and we set the value of $f_n(C)$ to zero for $K(C)<n\leq K_0$. The coefficient $c_n(C)$ then reads
\begin{equation}
c_n(C)=\frac{f_n(C)}{b_0}-\sum_{m=0}^{n-1}\frac{b_{n-m}}{b_0}c_{m}(C),
\label{cn}
\end{equation}
for $0<n\leq K(C)$ and
\begin{equation}
c_n(C)=-\sum_{m=1}^{\textrm{min}\{K_0,n\}}\frac{b_{m}}{b_0}c_{n-m}(C),
\label{cn_gt_K}
\end{equation}
for $n>K(C)$. Details of this calculation are presented in Appendix \ref{appendix_a}. 

Furthermore, we can use the fact that $\sum_{C}P(C)=1$, which translates to the following condition on the coefficients $c_n(C)$ for any $n$,
\begin{equation}
\sum_{C}c_n(C)=\delta_{n,0},\quad n=0,\dots,K_0,
\label{cn_sum}
\end{equation}
where $\delta_{n,0}$ is the Kronecker delta function.

In this paper we compute low-order coefficients for the Taylor expansion of $P(C)$ in $\kappa=\alpha\ll\beta,\omega_i$ (Section \ref{small_alpha}) and $\kappa=\beta\ll\alpha,\omega_i$ (Section \ref{small_beta}). It is important to emphasise that these Taylor coefficients are exact. Thus the only approximation that we make is to replace the Taylor series with the corresponding Taylor polynomial, which is valid when the value of the expansion parameter is small compared to other transition rates.  

\subsection{Power series in $\alpha$}
\label{small_alpha}

\subsubsection{Zero-order coefficients}

As it can be seen from Eq.~(\ref{e0}), the only configuration for which $e_0(C)=0$ when $\kappa=\alpha$ is the empty lattice, which we denote by $C=\emptyset$. Following the argument given in Sec.~\ref{main_idea}, Eq.~(\ref{master_0}) applies for all configurations except the empty one and it has the trivial solution $f_0(C)=0$ if $C \neq \emptyset$ . We thus conclude that the value of $f_0(C)$ is non-zero only for the empty configuration. Therefore, we can write $f_0(C)=\delta_{C,\emptyset}f_0(\emptyset)$ and inserting it into Eq.~(\ref{c0}) yields
\begin{equation}
c_0(C)=\frac{f_0(C)}{\sum_{C}f_0(C)}=\frac{\delta_{C,\emptyset}f_0(\emptyset)}{\sum_C \delta_{C,\emptyset} f_0(C)}=\delta_{C,\emptyset} \,,
\label{c0_alpha}
\end{equation}
which after insertion into (\ref{Pseries}) gives
\begin{equation}
P(C)=\delta_{C,\emptyset}+O(\alpha).
\label{P0}
\end{equation}
The zeroth-order solution is therefore equivalent to setting $\alpha=0$ in the original master equation: when particles are not allowed to enter, the steady state is an empty lattice.

\subsubsection{Proof that many coefficients $f_n(C)$ are equal to zero}

We can use the Schnakenberg network theory to show that many coefficients $f_n(C)$ are zero too, not just for $n=0$. Let us consider a configuration $C\neq\emptyset$ and let us define $N(C)$ as the number of particles in $C$
\begin{equation}
N(C)=\sum_{i=1}^{L}\tau_{i}(C).
\end{equation}
According to the Schnakenberg network theory, 
\begin{equation}
P(C)\propto \sum_{n=0}^{K(C)}f_n(C)\alpha^n=\sum_{T_C}w(T_C),
\label{SNT}
\end{equation}
where the last sum runs over all spanning trees $T_C$ rooted at $C$ and $w(T_C)$ is the product of the rates of all transitions (edges) contained in $T_C$. For a given $T_C$, there is a unique path from \emph{any} $C'\neq C$ to $C$. Let us now consider $C'$ to be the empty configuration, $C'= \emptyset$. A directed path from $C'=\emptyset$ to $C$ must include \emph{at least} $N(C)$ transitions at rate $\alpha$, since a number $N(C)$ of particles must enter the lattice in order to reach configuration $C$ (note that we say at least because some particles may as well leave the system). In order for the relation in Eq.~(\ref{SNT}) to hold, we
thus conclude that $f_n(C)=0$ if $n$ is smaller than the number of particles $N(C)$ present in configuration $C$, or equivalently,
\begin{equation}
\label{zerof}
f_n(C) \neq 0\,\quad\textrm{if and only if}\quad N(C) \le n\le K(C).
\end{equation}

Relation (\ref{zerof}) is crucial for our analysis, as it drastically reduces the number of unknowns. For $n=0$, it shows that $f_0(C)\neq 0$ for only one configuration, the empty lattice. For $n=1$, there are $L+1$ configurations that have non-zero coefficients, $L$ configurations with one particle and the empty lattice, where $L$ is the number of sites in the lattice. For $n=2$, the number of non-zero coefficients is $L(L-1)/2+L+1$, where $L(L-1)/2$ is the number of configurations with two particles, and $L+1$ is the number of configurations with 1 or 0 particles. In general, the number of non-zero coefficients of order $n$ grows polynomially in $L$ and is given by the partial sum of binomial coefficients, $\sum_{j=0}^{n}\binom{L}{j}$.

\subsubsection{First-order coefficients}
\label{first-order}

According to relation~(\ref{zerof}), for $n=1$, we need to consider configurations with at most one particle. Let us label a configuration with one particle at site $i$ by $1_i$. The equations for $f_1(1_i)$, $i=1,\dots,L$ are given by
\begin{subequations}
	\begin{align}
	\omega_1 f_1(1_1)&=f_0(\emptyset),\label{f1_1}\\
	\omega_i f_1(1_i)&=\omega_{i-1}f_1(1_{i-1}),\quad i=2,L-1,\label{f1_i}\\
	\beta f_1(1_L)&=\omega_{L-1}f_1(1_{L-1}),\label{f1_L}
	\end{align}
\end{subequations}
where we have set $f_1(C)=0$ for all $C$ that contain more than one particle, following relation~(\ref{zerof}). These equations can be easily solved, yielding
\begin{equation}
\label{f1_solution}
f_1(1_i)=\frac{f_0(\emptyset)}{\omega_i},
\end{equation}
where we have introduced the notation $\omega_L=\beta$. Inserting Eq. (\ref{f1_solution}) into (\ref{bn}) and (\ref{cn}) gives
\begin{subequations}
	\begin{align}
	& c_1(\emptyset)=-\sum_{i=1}^{L}\frac{1}{\omega_i},\label{c1_0}\\
	& c_1(1_i)=\frac{1}{\omega_i},\quad i=1,\dots,L\label{c1_i}\\
	& c_1(C)=0, \quad 2\leq\sum_{i=1}^{L}\tau_i(C)\leq L.
	\end{align}
\end{subequations}

The average particle current $J$ is obtained by inserting Eqs.~(\ref{c0_alpha}), (\ref{c1_0}) and (\ref{c1_i}) into Eq.~(\ref{Pseries}) and then into (\ref{current}), which up to the quadratic term yields
\begin{equation}
\label{J_2}
J=\alpha-\frac{1}{\omega_1}\alpha^2+O(\alpha^3).
\end{equation}
Similarly, the expressions for the local and total density $\rho_i$ and $\rho$ read, respectively,
\begin{align}
& \rho_i=\frac{\alpha}{\omega_i}+O(\alpha^2),\label{rho1_i}\\
& \rho=\frac{1}{L}\left(\sum_{i=1}^{L}\frac{1}{\omega_i}\right)\alpha+O(\alpha^2).\label{rho1}
\end{align}
We can check that Eqs.~(\ref{J_2}) and (\ref{rho1}) give the familiar result for the homogeneous case with $\omega_i = \omega$, $J=\alpha(1-\alpha/\omega)$ and $\rho=\alpha/\omega$.

\subsubsection{Second-order coefficients}
\label{second-order}

According to relation~(\ref{zerof}), for $n=2$ we need to consider configurations with at most two particles. As before, we label configurations by the positions of their particles, so that $C=1_i 1_j$ for $i=1,\dots,L-1$ and $j=i+1,\dots,L$ denote configurations with two particles and $C=1_i$ for $i=1,\dots,L$  denote configurations with one particle. 

We first look at configurations with a particle at site $1$. Essentially, we have to consider three main cases: $C=1_1 1_2$, $C=1_1 1_j$ with $j>2$, and $C=1_1$. Applying Eq.~(\ref{master_n}) to these three cases yields
\begin{subequations}
	\begin{align}
	& \omega_2 f_2(1_1 1_2)=f_1(1_2),\label{f2_12}\\
	& (\omega_1+\omega_j)f_2(1_1 1_j)=f_1(1_j)+\omega_{j-1}f_2(1_1 1_{j-1}),\label{f2_1j}\\
	&\omega_1 f_2(1_1)=f_1(\emptyset)+\omega_L f_2(1_1 1_L),\label{f2_1}
	\end{align}
\end{subequations}
where $f_1(1_j)=f_0(\emptyset)/\omega_j$. Eq.~(\ref{f2_1j}) is a recurrence relation in $j$ with Eq.~(\ref{f2_12}) as a initial condition. This is a non-homogeneous recurrence relation with variable coefficients that can be turned into a non-homogeneous recurrence relation with constant coefficients and solved explicitly, yielding

\begin{subequations}
	\begin{align}
	& f_2(1_1 1_2)=\frac{f_0(\emptyset)}{\omega_{2}^{2}},\label{f2_12_final}\\
	& f_2(1_1 1_j)=\frac{f_0(\emptyset)}{\omega_1 \omega_j}\left[1+\left(\frac{\omega_1}{\omega_2}-1\right)\prod_{q=3}^{j}\frac{\omega_q}{\omega_1+\omega_q}\right].\label{f2_1j_final}
	\end{align}
\end{subequations}
We can now compute $f_2(1_1)$ by inserting Eq.~(\ref{f2_1j_final}) for $j=L$ into Eq.~(\ref{f2_1}).
%
%
Inserting Eqs.~(\ref{f2_1}), (\ref{f2_12_final}) and (\ref{f2_1j_final}) into (\ref{cn}) gives the following expressions for $c_2(1_1 1_j)$ and $c_1(1_1)$, respectively, 
\begin{subequations}
	\begin{align}
	& c_2(1_1 1_2)=\frac{1}{\omega_{2}^{2}},\label{c2_12}\\
	& c_2(1_1 1_j)=\frac{1}{\omega_1 \omega_j}\left[1+\left(\frac{\omega_1}{\omega_2}-1\right)\prod_{q=3}^{j}\frac{\omega_q}{\omega_1+\omega_q}\right],\label{c2_1j}\\
	& c_2(1_1)=\frac{1}{\omega_{1}^{2}} \left[1+\left(\frac{\omega_1}{\omega_2}-1\right)\prod_{q=3}^{L}\frac{\omega_q}{\omega_1+\omega_q}\right]\nonumber\\
	&\qquad-\frac{1}{\omega_1}\sum_{i=1}^{L}\frac{1}{\omega_i}.\label{c2_1}
	\end{align}
\end{subequations}
It is important to note that these coefficients do not depend on the coefficients $f_0(\emptyset)$ and $f_1(\emptyset)$. Otherwise, our method would not be useful as we cannot say anything about these coefficients \footnote{In Ref.~\cite{SzavitsNossan13} it was erroneously stated that all $f_n(\emptyset)$ are equal.}. We show in Section~\ref{fn_f0_expression} that $f_n(\emptyset)$ are not needed until we want to compute $c_n(C)$ for $n>K(C)$, which is equivalent of solving the full master equation (\ref{master}).

We argue that these are the only second-order coefficients that we need in order to compute the cubic term in the power series of $J$, since
\begin{align}
J &=\alpha\sum_{C}P(C)\left[1-\tau_1(C)\right]=\sum_{n=0}^{\infty}\left(\sum_{\substack{C\\ \tau_1=0}}c_n(C)\right)\alpha^{n+1}\nonumber\\
&=\sum_{n=0}^{\infty}\left[\delta_{n,0}-\sum_{\substack{C\\ \tau_1=1}}c_n(C)\right]\alpha^{n+1}=\alpha-\frac{1}{\omega_1}\alpha^2\nonumber\\
&\quad-\left[\sum_{j=2}^{L}c_2(1_1 1_j)+c_2(1_1)\right]\alpha^{3}+O(\alpha^4),\label{J_expansion}
\end{align}
where we have used Eq.~(\ref{cn_sum}) in the second line. After inserting (\ref{c2_12})-(\ref{c2_1}) into (\ref{J_expansion}), the expression for $J$ reads
\begin{align}
J & =\alpha-\frac{1}{\omega_1}\alpha^2+\bigg(\frac{1}{\omega_1}-\frac{1}{\omega_2}\bigg)\Bigg[\frac{1}{\omega_2}+\sum_{j=3}^{L}\bigg(\frac{1}{\omega_j}\nonumber\\
& +\delta_{j,L}\frac{1}{\omega_L}\bigg)\prod_{q=3}^{j}\frac{\omega_q}{\omega_1+\omega_q}\Bigg]\alpha^3+O(\alpha^4).\label{J_3}
\end{align}
We note that the cubic term in Eq.~(\ref{J_3}) is equal to zero when $\omega_1=\omega_2$. This includes the homogeneous case in which $\omega_1=\dots=\omega_{L-1}=1$, which was known before from the exact solution~\cite{DEHP}.

\subsubsection{Power series solution in the mean-field approximation}

Before we compare our predictions to exact (numerical) results from Monte Carlo simulations in Section~\ref{MC}, it is instructive to analyse how Eq.~(\ref{J_3}) compares to the mean-field (MF) solution used in other approaches~\cite{Shaw04,Reuveni11}. 

The MF approximation amounts to replacing $\langle\tau_i\tau_j\rangle$ for $i\neq j$ with $\rho_i\rho_j$, where $\rho_i=\langle\tau_i\rangle$ is the local particle density. This approximation leads to the following mean-field equations for $J$ and $\rho_i$
\begin{eqnarray}
\label{mf}
J &=& \alpha(1-\rho_1)=\omega_1\rho_1(1-\rho_2)=\dots\nonumber\\
  &=& \omega_{L-1}\rho_{L-1}(1-\rho_L)=\beta\rho_L.
\end{eqnarray}
The exact (closed-form) solution of these equations is unknown. Instead, we can look for a perturbative solution for small $\alpha$ in the following form
\begin{equation}
\label{rhoi_mf}
\rho_i=\sum_{n=0}^{\infty}\rho_{i}^{(n)}\alpha^n.
\end{equation}
Inserting Eq.~(\ref{rhoi_mf}) into (\ref{mf}) and collecting terms of the same power of $\alpha$ for small $n$ leads to
\begin{eqnarray}
\label{rhoi_mf2}
\rho_{i}^{(0)} &=& 0,\quad \rho_{i}^{(1)}=\frac{1}{\omega_i},\nonumber\\
\rho_{i}^{(2)} &=& \begin{cases}\frac{1}{\omega_i}\left(\frac{1}{\omega_{i+1}}-\frac{1}{\omega_1}\right), & i=1,\dots,L-1\\ -\frac{1}{\omega_{1}\omega_L}, & i=L.\end{cases}
\end{eqnarray}
Inserting these coefficients back into Eq.~(\ref{rhoi_mf}) and then into (\ref{mf}), we get the following expression for current $J_{MF}$ in the mean-field approximation
\begin{equation}
\label{J_MF}
J_{MF}=\alpha-\frac{1}{\omega_1}\alpha^2+\left(\frac{1}{\omega_1}-\frac{1}{\omega_2}\right)\frac{1}{\omega_1}\alpha^3+O(\alpha^4).
\end{equation}
While the first two terms in Eqs.~(\ref{J_3}) and (\ref{J_MF}) are the same, the third term in Eq.~(\ref{J_MF})  depends only on $\omega_1$ and $\omega_2$, which is markedly different from the third term in Eq.~(\ref{J_3}), which depends on all $\omega_i$. Hence, the mean-field results clearly deviate from the main result of this paper, i.e., the power series solution derived in Eq.~(\ref{J_3}).

\subsubsection{Monte Carlo simulations}
\label{MC}

We now compare the power series of the current $J(\alpha)$ to the exact current that we compute numerically using kinetic Monte Carlo simulations based on the Gillespie algorithm for $L=250$ lattice sites. For that purpose we generated $\omega_1,\dots,\omega_L$ randomly from the uniform distribution on $[1,10]$, which are plotted in Fig.~\ref{fig4}. 

\begin{figure}[hbt]
	\includegraphics[width=8cm]{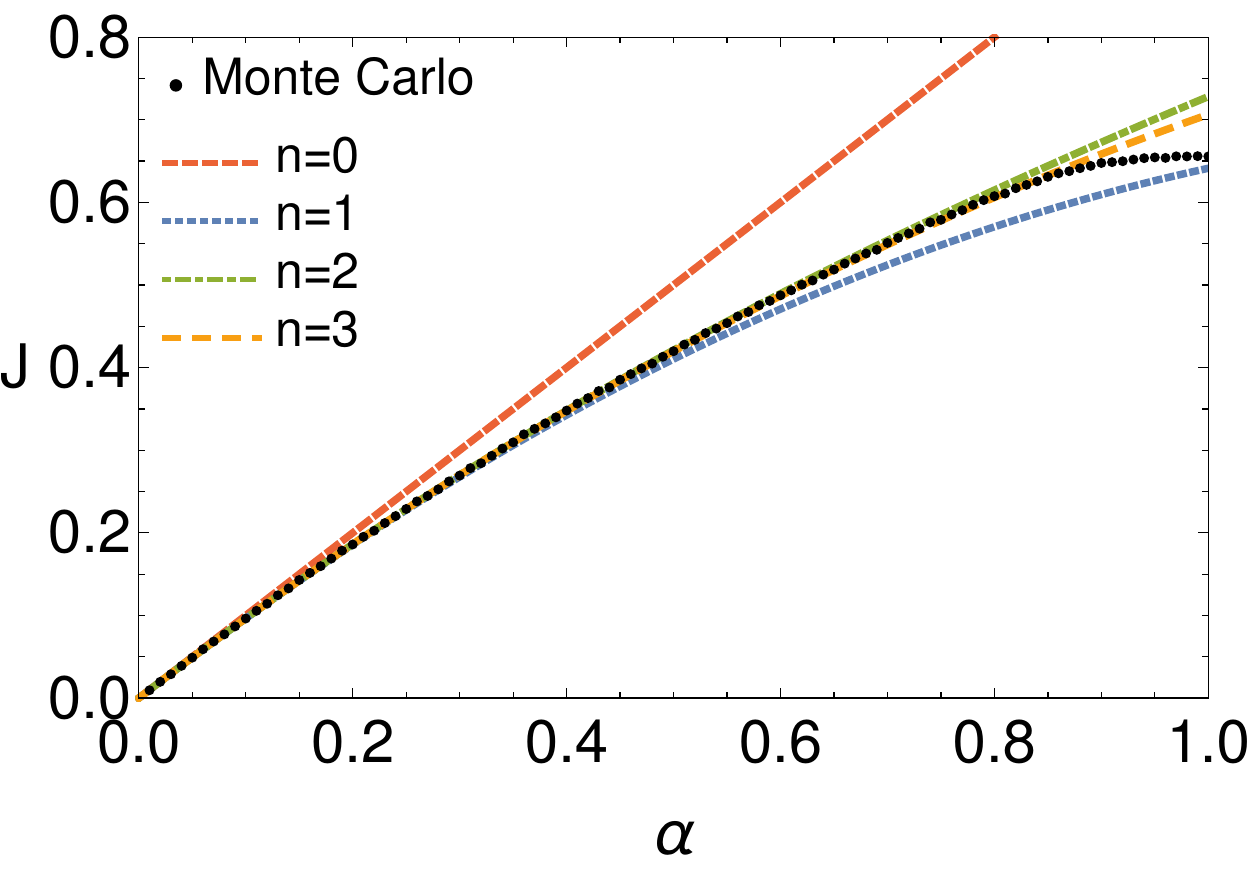}
	\caption{Current $J$ as a function of the entrance rate $\alpha$. The black dots show the result of simulations obtained using the Gillespie algorithm (averaged over $500$ independent runs). The dashed red ($n=0$) and dotted blue ($n=1$) lines are linear and quadratic approximations from Eq.~(\ref{J_2}), respectively. The dot-dashed green ($n=2$) line is cubic approximation from Eq.~(\ref{J_3}). The dashed orange ($n=3$) line is quartic approximation, which was computed numerically.\label{fig3}}
\end{figure}

In Fig.~\ref{fig3}, the first three terms in the power series of $J$ in $\alpha$ corresponding to $n=0,1$ and $2$ were computed analytically, using Eqs.~(\ref{J_2}) and~(\ref{J_3}), respectively. The fourth term corresponding to $n=3$ was computed numerically using a simple algorithm that we present in Appendix~\ref{appendix_b}. The agreement between the simulation results and the analytical expressions visibly increases as we include more terms in the power series.

\begin{figure}[hbt]
	\centering\includegraphics[width=8cm]{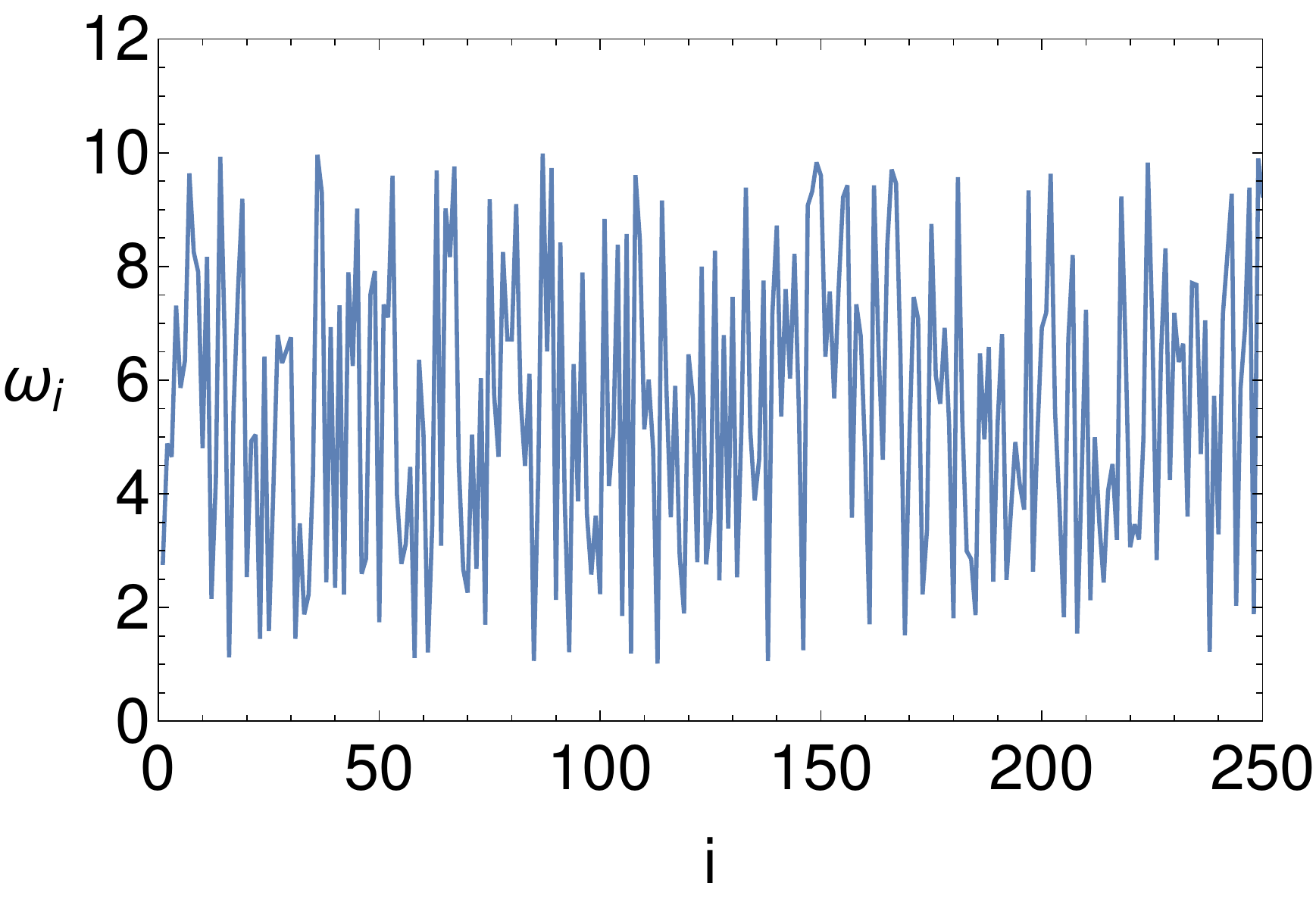}
	\caption{Hopping rates $\omega_i$ for $i=1,\dots,L=250$ used to compute $J$ in Fig.~\ref{fig3}. \label{fig4}}
\end{figure}

\begin{figure}[hbt]
	\includegraphics[width=8cm]{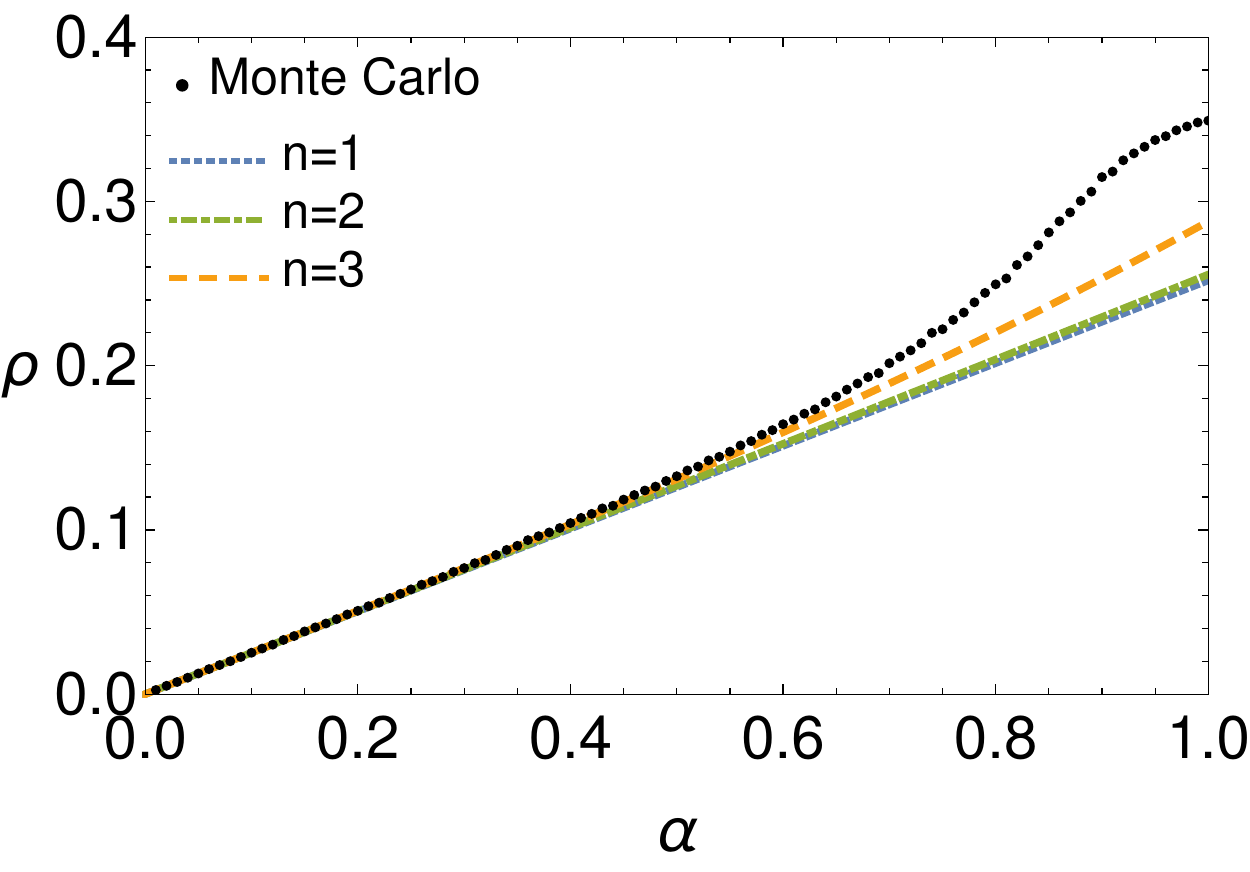}
	\caption{Total density $\rho$ as a function of the entrance rate $\alpha$. The black dots show the result of simulations obtained using the Gillespie algorithm (averaged over $500$ independent runs). The dashed blue line ($n=1$) is linear approximation from Eq.~(\ref{rho1}). The dotted green ($n=2$) and dashed orange ($n=3$) lines are quadratic and cubic approximations, respectively, which were computed numerically.\label{fig5}}
\end{figure}

Unlike $J$, the total density $\rho$ depends on all second-order coefficients $c_2(C)$. The equation for $f_2(1_i 1_j)$ is a recurrence relation in two indices $i$ and $j$, which is difficult to solve explicitly. In Fig.~\ref{fig5} we compare the power series of the total density $\rho(\alpha)$ for the same choice of $\omega_i$ as in Fig.~\ref{fig3} with the Monte Carlo simulations. The linear term corresponding to $n=1$ was computed from Eq.~(\ref{rho1}), while the quadratic and cubic approximations corresponding to $n=2$ and $3$ were computed numerically.

As we can already see for the second-order coefficients, handling higher orders become progressively more difficult. In the next two Sections we develop a general method for computing higher-order coefficients $c_n(C)$, revealing an interesting connection with combinatorial objects called Young tableaux.

\subsubsection{Higher-order coefficients}
\label{fn_f0_expression}

Our starting point is recurrence relation in Eq.~(\ref{master_n}), which after rearranging gives
\begin{align}
& f_n(C)=\sum_{C'}\frac{I_{C',C}}{e_{0}(C)}f_{n-1}(C')-\left(\sum_{C'}\frac{I_{C,C'}}{e_{0}(C)}\right)f_{n-1}(C)\nonumber\\
&+\sum_{C'}\frac{(1-I_{C',C})}{e_{0}(C)}W(C'\rightarrow C)f_n(C'),\quad C\neq\emptyset.
\label{recurrence}
\end{align}
We cannot immediately find $f_n(C)$ by iterating over $n$, because the r.h.s. of Eq.~(\ref{recurrence}) contains coefficients of order $n$ as well. The solution is to picture Eq.~(\ref{recurrence}) as a recurrence relation not only in order $n$, but also in the configuration $C$. The iteration stops when all coefficients on the r.h.s. of Eq.~(\ref{recurrence}) belong to the empty configuration $C'=\emptyset$. That is because the recurrence relation in Eq.~(\ref{recurrence}) does not hold for $C=\emptyset$. The final expression for $f_n(C)$ for $N(C)\leq n\leq K(C)$ must then be of the form
\begin{equation}
f_n(C)=\sum_{m=N(C)}^{n}\mu_{m}(C)f_{n-m}(\emptyset), 
\label{fn_f0}
\end{equation}
where $\mu_m(C)$ is an unknown coefficient. As we show later in more detail, the sum starts from $m=N(C)$ because of Eq.~(\ref{zerof}).

The expression for $f_n(C)$ in Eq.~(\ref{fn_f0}) can be used to calculate $c_n(C)$ in the power series of $P(C)$. According to Eqs.~(\ref{bn}) and (\ref{cn}), the expression for $f_n(C)$ for 
$N(C)\leq n\leq K(C)$ is given by
\begin{equation}
f_n(C)= b_0 c_n(C) +\sum_{m=0}^{n-1}b_{n-m} c_m (C)\;.
\label{fn_cn}
\end{equation}
By inserting this expression for $C=\emptyset$ into Eq.~(\ref{fn_f0}) and comparing the term containing $b_0$ to the one in Eq.~(\ref{fn_cn}), we conclude that
\begin{equation}
c_n(C)=\sum_{m=N(C)}^{n}\mu_m(C)c_{n-m}(\emptyset).
\label{cn_c0}
\end{equation}
The recurrence relation for $c_n(\emptyset)$ can be obtained by inserting Eq.~(\ref{cn_c0}) into Eq.~(\ref{cn_sum}), which gives
\begin{equation}
c_n(\emptyset)=-\sum_{C\neq\emptyset}\sum_{m=N(C)}^{n}\mu_m(C)c_{n-m}(\emptyset),
\label{c0_recurrence}
\end{equation}
and the initial condition is $c_0(\emptyset)=1$.

The expression for $c_n(C)$ in Eq.~(\ref{cn_c0}), together with the recurrence relation for $c_n(\emptyset)$ in Eq.~(\ref{c0_recurrence}) is one the main results of this paper. It shows that the first $K(C)$ coefficients in the power series of $P(C)$ in $\alpha$ are determined by the coefficients $\mu_m(C)$ and not by the unknown coefficients $f_{m}(\emptyset)$ that our method cannot determine. The situation changes for $n>K(C)$, for which Eq.~(\ref{fn_cn}) is replaced by
\begin{equation}
c_n(C)=\sum_{m=1}^{\textrm{min}\{K_0,n\}}\frac{b_m}{b_0}c_{n-m}(C),\quad n>K(C),
\label{cn_bn}
\end{equation}
where we remind that $K_0=\textrm{max}_C\{K(C)\}$. In this case $c_n(C)$ depends on $b_m$, which in turn depends on the unknown coefficients $f_{m}(\emptyset)$. In other words, we cannot find $c_n(C)$ without finding all $f_m(\emptyset)$. That is not surprising: according to Eqs.~(\ref{f}) and (\ref{fn_f0}), finding all $\mu_m(C)$ and $f_m(\emptyset)$ amounts to find the full solution of the master equation. In that context, the result in Eq.~(\ref{cn_bn}) tells us that we cannot say anything about $c_n(C)$ for $n>K(C)$ without solving the original master equation fully. Since we expect $K(C)$ to grow exponentially with the system size $L$, these coefficients are inaccessible for all practical purposes. 

In the next two Sections we complete the power series of $P(C)$ in $\alpha$ by showing how to compute the coefficients $\mu_m(C)$.

\subsubsection{Back-substitution method for finding $\mu_n(C)$}
\label{mu}
In this Section we show how to find $f_n(C)$ and thus $\mu_n(C)$ by backward substitution until all terms on the r.h.s. of Eq.~(\ref{recurrence}) belong to the empty configuration $\emptyset$.  

The first sum on the r.h.s. of Eq.~(\ref{recurrence}) is zero unless $C$ contains a particle at site $1$,
\begin{equation}
\sum_{C'}\frac{I_{C',C}}{e_0(C)}f_{n-1}(C')=\frac{\tau_i(C)}{e_0(C)}f_{n-1}(0,\tau_2,\dots).
\end{equation}
This contribution to $f_n(C)$ is obtained from $C$ by removing a particle at site $1$, provided $\tau_1(C)=1$. The resulting coefficient is $f_{n-1}(0,\tau_2(C),\dots,\tau_L(C))$, which is multiplied by the weight $1/e_0(C)$. We call this step {\it rule $1$}.

The second sum is zero unless site $1$ in $C$ is empty
\begin{equation}
\left(\sum_{C'}\frac{I_{C,C'}}{e_0(C)}\right)f_{n-1}(C)=\frac{1-\tau_i(C)}{e_0(C)}f_{n-1}(C).
\end{equation}
In this step we reduce the order of $f_n(C)$ to $f_{n-1}(C)$, provided $\tau_1(C)=0$. The resulting coefficient $f_{n-1}(C)$ is multiplied by the weight $(-1)/e_{0}(C)$. We call this step {\it rule $2$}.

The third sum runs over all configurations $C'$ that lead to $C$ by moving one particle forwards. Each of these moves contributes to $f_n(C)$ with $f_n(C')$, multiplied by the weight $W(C'\rightarrow C)/e_0(C)$. We call this step {\it rule $3$} if a pair of variables  ($\tau_i=0$,$\tau_{i+1}=1$) in $C$ is replaced by ($\tau_{i}=1$,$\tau_{i+1}=0$) in $C'$ and {\it rule $4$} if $\tau_{L}=0$ in $C$ is replaced by $\tau_{L}=1$ in $C'$. 

\begin{table}[hbt]
	\caption{\label{table1}Iteration steps (rules) for going from $C'$ on the r.h.s. to $C$ on the l.h.s. of Eq.~(\ref{recurrence}).}
	\begin{ruledtabular}
		\begin{tabular}{llll}
			Rule & Iteration ($C'\rightarrow C$) & Coefficient & Weight\\
			\hline
			$1$ & $0_1\rightarrow 1_1$ & $f_{n-1}(C')$ & $1/e_0(C)$\\
			$2$ & $0_1\rightarrow 0_1$ & $f_{n-1}(C)$ & $(-1)/e_0(C)$\\
			$3$ & $1_i 0_{i+1}\rightarrow 0_i 1_{i+1}$ & $f_n(C')$ & $\omega_{i}/e_0(C)$\\
			$4$ & $1_L\rightarrow 0_L$ & $f_n(C')$ & $\beta/e_0(C)$
		\end{tabular}
	\end{ruledtabular}
\end{table}

In summary, rule $1$ reduces order $n$ to $n-1$ and removes a particle from site $1$. Rule $2$ reduces order $n$ to $n-1$ but leaves the configuration unchanged. Rule $3$ moves a particle at site $2\leq i\leq L$ to $i-1$, provided site $i-1$ is empty. Rule $4$ moves a particle from the right reservoir to site $L$, provided site $L$ is empty. A summary of all possible transitions is presented in Table \ref{table1}.

The idea is to repeat these rules for all coefficients on the r.h.s. of Eq.~(\ref{recurrence}) until we get a coefficient $f_m(\emptyset)$ that belongs to the empty configuration. Since the recurrence relation in Eq.~(\ref{recurrence}) does not apply to $C=\emptyset$, we leave this coefficient as it is. The other possibility is to get a coefficient $f_m(C')$ for which the number of particles $N(C')>m$. This can happen because of the rule $4$ that increases the number of particles. However, this coefficient does not contribute to $f_n(C)$ because of Eq.~(\ref{zerof}). This is how we obtain the solution for $f_n(C)$ presented in Eq.~(\ref{fn_f0}).

\textit{Example}. Let us say we want to compute $f_2(1_1 1_3)$ for a lattice of $L=4$ sites. We can apply  rule $1$ to $C=1_1 1_3$ leading to $C'=1_3$ and rule $3$ leading to $C'=1_1 1_2$. The corresponding weights are $1/e_0(1_1 1_3)$ and $\omega_2/e_0(1_1 1_2)$, respectively. On this occasion we do not apply rule $4$ leading to $C'=1_1 1_3 1_4$ because we know from Eq.~(\ref{zerof}) that $f_2(1_1 1_3 1_4)=0$. We can thus write $f_2(1_1 1_3)$ as
\begin{equation}
f_2(1_1 1_3)=\frac{1}{e_0(1_1 1_3)}f_1(1_3)+\frac{\omega_2}{e_0(1_1 1_3)}f_2(1_1 1_2).
\end{equation}
Now we look at each of the configurations $1_3$ and $1_1 1_2$ on the r.h.s. separately. We apply rule $1$ to $C=1_3$ leading to $C'=1_2$, whereby the corresponding weight is $\omega_2/e_0(1_3)$. Similarly, we apply rule $1$ to $C=1_1 1_2$ leading to $C'=1_2$, whereby the corresponding weight is $1/e_0(1_2)$. Now $f_2(1_1 1_3)$ reads
\begin{eqnarray}
f_2(1_1 1_3)&=&\frac{\omega_2}{e_0(1_1 1_3)e_0(1_3)}f_1(1_2)\nonumber\\
&+&\frac{\omega_2}{e_0(1_1 1_3)e_0(1_1 1_2)}f_1(1_2).
\end{eqnarray}
Next, we apply rule $3$ to $C=1_2$ leading to $C'=1_1$, which is weighted by $\omega_1/e_0(1_2)$. In the final step we apply rule $1$ to $C=1_1$ leading to $C'=\emptyset$, which is weighted by $1/e_0(1_1)$. Altogether, the expression for $f_2(1_1 1_3)$ reads
\begin{eqnarray}
f_2(1_1 1_3)&=&\frac{\omega_1\omega_2}{e_0(1_1 1_3)e_0(1_3)e_0(1_2)e_0(1_1)}f_0(\emptyset)\nonumber\\
&+&\frac{\omega_1\omega_2}{e_0(1_1 1_3)e_0(1_1 1_2)e_0(1_2)e_0(1_1)}f_0(\emptyset),
\end{eqnarray}
so that $\mu_2(1_1 1_3)$ is given by
\begin{eqnarray}
\mu_2(1_1 1_3)&=&\frac{\omega_1\omega_2}{e_0(1_1 1_3)e_0(1_3)e_0(1_2)e_0(1_1)}\nonumber\\
&+&\frac{\omega_1\omega_2}{e_0(1_1 1_3)e_0(1_1 1_2)e_0(1_2)e_0(1_1)}.
\end{eqnarray}

In the next Section we provide a formal expression for the coefficients $\mu_m(C)$. In particular, we show that the successive application of rules 1--4 can be graphically represented by  combinatorial objects called Young tableaux of shifted shape.

\subsubsection{Coefficients $\mu_m$ and Young tableaux}

By definition, $\mu_m(C)$ is a sum of products of weights of all iteration steps in Table~\ref{table1} that connect $f_n(C)$ to $f_{n-m}(\emptyset)$ in Eq.~(\ref{fn_f0}). By construction, we know immediately that 
\begin{equation}
\mu_m(C)=0,\quad m<N,
\end{equation}
where $N$ is the number of particles in $C$. We thus have to consider only $m\geq N$.

Let us first consider the case of $m=N$ for which $f_{N}(C)=\lambda_N(C)f_0(\emptyset)$. In order to compute $\mu_N(C)$, we have to find all directed paths from $\emptyset$ to $C$ that insert exactly $N$ particles from the left reservoir. Let us label the $N$ particles by integers $1,2\dots,N$ in the order in which they enter the lattice. Next, we represent a directed path from $\emptyset$ to $C$ by a sequence of moves, whereby each move in the sequence is labelled by the label of the particle that made that move. 

For example, in order to get from $\emptyset$ to $1_2 1_4$, we need to construct integer sequences consisting of four labels $1$ and two labels $2$. In the first move, we insert particle $1$ from the left reservoir and the resulting sequence is $1$. In the second move, we move particle $1$ again and the resulting sequence is $11$. In the third move, we can either move particle $1$ leading to $111$ or we can insert particle $2$ into the lattice leading to $112$. We repeat these steps until particle $1$ reaches site $4$ and particle $2$ reaches site $2$. The list of all final sequences representing directed paths from $\emptyset$ to $1_2 1_4$ is $111122$, $111212$, $111221$, $112112$, $112121$, five in total.

A distinguished property of these integer sequences is that the number of $1$'s is strictly greater than the number of $2$'s. This is not only true for the final sequence, but also for all initial sub-sequences leading to the final sequence and is a direct consequence of exclusion that forbids two or more particles to share the same lattice site. For example, sequence $121112$ in the example above is forbidden because its initial sub-sequence is $12$, which implies that particle $2$ moves onto site $1$ that is already occupied by particle $1$. The same property extends to sequences and all their initial sub-sequences that contain more than two particles, so that the number of particles labelled by $k$ is strictly larger than the number of particles labelled by $k+1$, for all $k=1,\dots,N-1$. 

Sequences of integers $1$ and $2$ in which the number of $1$ remains greater than the number of integers $2$ in all initial sub-sequences have a long history in mathematics (see Ref.~\cite{ballot} for example). The famous Bertrand's Ballot Problem asks for the probability that candidate $A$ receiving $p$ votes in total stays ahead of candidate $B$ receiving $q<p$ votes in total as the votes are counted~\cite{Feller68}. When there are more then two candidates, we speak of generalised ballot sequences. Confusingly, the names \emph{generalised ballot sequence}, \emph{lattice permutation} or \emph{lattice word} that are nowadays used for such sequences assume that the number of integers $k$ is \emph{no less} than the number of integer $k+1$, rather than being strictly greater. Terms ``weak'' in the former and ``strict'' in the latter case are sometimes used to distinguish these two cases.

There is a graphical way to represent generalised ballot sequences using combinatorial objects called Young tableaux~\cite{Adin15}. A \emph{Young diagram} of shape $\lambda=(\lambda_1,\dots,\lambda_k)$ is a left-justified shape of $k$ rows of boxes of length $\lambda_1,\dots,\lambda_k$ \footnote{We are using here the English notation for the appearance of Young diagrams, see Ref.~\cite{Adin15} for other notations.}. A Young diagram whose boxes are filled with integers $1,2,\dots,\lambda_1+\dots+\lambda_k$ that are strictly increasing along rows and columns is called a \emph{standard} Young tableau. A Young tableau of shifted shape $\lambda=(\lambda_1,\dots,\lambda_k)$ is a standard Young tableau whose $i$-th row is indented by $i-1$ boxes. In the following example, 
\begin{equation}
\yng(4,2),\quad\young(1256,34)\quad\textrm{and}\quad\young(1235,:46),
\label{syt_example}
\end{equation}
the first is a Young diagram of shape $(4,2)$, the second is a standard Young tableau of shape $(4,2)$ and the third is a standard Young tableau of shifted shape $(4,2)$. A standard Young tableau corresponds to a weak generalised ballot sequence and a standard Young tableau of shifted shape corresponds to a strict generalised ballot sequence.

For a given configuration $C=1_{x(N)}\dots 1_{x(1)}$ with $N$ particles at positions $x(1)>x(2)>\dots>x(N)$ (labelled in the order in which they enter the lattice), there is a one-to-one correspondence between a directed path from $\emptyset$ to $C$ and a Young tableau of shifted shape $x=(x(1),\dots,x(N))$. We recall that a directed path $S(C)$ from $\emptyset$ to $C$ is a sequence of $x(1)$ integers $1$, $x(2)$ integers $2$ and so on, whereby the number of integers $k$ is strictly larger than the number of integers $k+1$ for all $k=1,\dots,N-1$ and for all sub-sequences of $S(C)$. Hence $x(i)$ designates both the position on the lattice of the particle $i$ (used to write the configuration $C$), and the number of moves that the particle has experienced in going from $\emptyset$ to $C$. The connection between $S(C)$ and the Young tableau of shifted shape $x$ is made by filling each box with a number $\in\{1,\dots,x(1)+\dots+x(N)\}$ that corresponds to the position of that move in the sequence. For example, a Young tableau of shifted shape $(4,2)$ in Eq.~(\ref{syt_example}) corresponds to a directed path from $\emptyset$ to $C=1_2 1_4$ that is represented by sequence $111212$ (Fig.\ref{fig3}).

\begin{figure}[hbt]
	\centering\includegraphics[width=6cm]{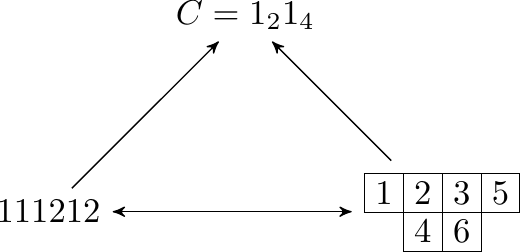}
	\caption{Correspondence between a directed path from $\emptyset$ to $C=1_{x(N)}\dots 1_{x(1)}$ and a Young tableaux of shifted shape $(x(1),\dots,x(N))$ for $N=2$, $x(1)=4$ and $x(2)=2$.}
	\label{fig6}
\end{figure}

We are now ready to state the expression for $\mu_N(C)$, where $N$ is the number of particles in $C=1_{x(N)}\dots 1_{x(1)}$. Let us denote by $t(x)$ a standard Young tableau of shifted shape $x=(x(1),\dots,x(N))$. This tableau corresponds to a sequence of $x(1)$ integers $1$, $x(2)$ integers $2$ and so, which we denote by $S(C)$. We call $C_k$ a configuration that corresponds to the initial sub-sequence of $S(C)$ of length $k$. The expression for $\mu_N(C)$ is then given by
\begin{equation}
\mu_N(1_{x(1)}\dots 1_{x(N)})=\prod_{i=1}^{N}\prod_{j=2}^{x(i)}\omega_{j-1}\sum_{t(x)}\prod_{k=1}^{l}\frac{1}{e_0(C_k)},
\label{mu_N}
\end{equation}
where the sum goes over all standard Young tableaux of shifted shape $x$. The product in front of the sum comes from the numerator of rule $3$ in Table \ref{table1} and depends on the corresponding Young \emph{diagram} but not on how the boxes are filled. The product after the sum goes over all iteration steps ($l$ in total) that lead from $\emptyset$ to $C$. For example, the expression for $\mu_2(1_2 1_4)$ is given by
\begin{align}
&\mu_2(1_2 1_4)=\omega_{1}^2\omega_2\omega_3\left[\frac{1}{\omega_1\omega_2\omega_3\omega_4(\omega_1+\omega_4)(\omega_2+\omega_4)}\right.\nonumber\\
&\quad+\left.\frac{1}{\omega_1\omega_2\omega_3(\omega_1+\omega_3)(\omega_1+\omega_4)(\omega_2+\omega_4)}\right.\nonumber\\
&\quad+\left.\frac{1}{\omega_1\omega_2\omega_3(\omega_1+\omega_3)(\omega_2+\omega_3)(\omega_2+\omega_4)}\right.\nonumber\\
&\quad+\left.\frac{1}{\omega_1\omega_{2}^2(\omega_1+\omega_3)(\omega_1+\omega_4)(\omega_2+\omega_4)}\right.\nonumber\\
&\quad+\left.\frac{1}{\omega_1\omega_{2}^{2}(\omega_1+\omega_3)\omega_3(\omega_2+\omega_4)}\right],
\end{align}
in which the terms in square brackets correspond respectively to sequences $111122$, $111212$, $111221$, $112112$ and $112121$.

So far we have discussed directed paths contributing to $\mu_m(C)$, where $m=N$ is the number of particles in $C$. The situation changes for $m>N$, because we may also apply rules $2$ and $4$ in addition to rules $1$ and $3$. Let us denote by $m_1$, $m_2$ and $m_4$ the number of times that we applied rules $1$, $2$ and $4$ in a directed path $S(C)$ from $\emptyset$ to $C$. Obviously, $m_1=N+m_4$ and $m=m_1+m_2$. 

Let us consider the case $m_2=0$ first, so that the number of particles entering and leaving the system is $m_1=m$ and $m_4=N-m$, respectively. We set $x(1)=\dots=x(m-N)=L+1$ for the particles that leave the system and the remaining particles have positions $x(m-N+1)>\dots>x(m)$. The only change for $m_4>0$ and $m_2=0$ compared to the previous case of $m_4=m_2=0$ (i.e. $m=N$) is that a Young tableau $t(x)$ that corresponds to $S(C)$ now has a shifted shape $x=(x(1)=L+1,\dots,x(m-N)=L+1,x(m-N+1),\dots,x(m))$, i.e. the first $m_4$ rows are of length $L+1$. 

However, care must be taken to exclude directed paths from $\emptyset$ to $C$ that revisit the empty lattice configuration. Let us denote by $t_{k,j}(x)$ the value of the $j$-th box in the $k$-th row of $t(x)$ starting from top to bottom. The requirement that the empty configuration is not revisited translates to the condition that $t_{k,L+1}(x)\neq k(L+1)$ for all $k=1,\dots,m_4=m_1-N$. Another way of stating this condition is to require that none of the sub-tableaux of $t(x)$, which are obtained by retaining only the first $k=1,\dots,m-1$ rows of $t(x)$, are themselves \emph{standard} Young tableaux of shifted shape $(x(1),\dots,x(k))$.

\textit{Example}. Let us consider the coefficient $\mu_2(1_1)$ for $L=3$, so that $m=2$, $N=1$, $m_1=2$, $m_2=0$ and $m_4=1$. In order to get from $\emptyset$ to $C=1_1$, we need to insert two particles and remove one ending with $x(1)=L+1=4$ and $x(2)=1$. The corresponding Young diagram is
\begin{equation*}
\young(~~~~,:~)
\end{equation*}
which we need to fill with integers $1,\dots,x(1)+x(2)=5$ that increase in all rows and columns. There are three possible fillings, which are given by
\begin{equation*}
\young(1234,:5),\quad \young(1245,:3),\quad \young(1235,:4).
\end{equation*}
However, the first filling corresponds to the sequence $11112$, which revisits the empty configuration after four jumps. Hence only the last two tableaux contribute to $\mu_2(1_1)$. 

For $m_2>0$, we note that the rule $2$ applied to any configuration $C'\in S(C)$ (provided $\tau_i(C')=0$) changes only the weight of $S(C)$, but not the sequence itself. Let us denote by $\epsilon_k$ the number of times that we applied the rule $2$ while being in configuration $C_k\in S(C)$, so that $\Vert\epsilon\Vert=\epsilon_1+\dots+\epsilon_l=m_2$. Since rule $2$ changes neither $S(C)$ nor the corresponding Young tableau, we call $\epsilon_k$ a \emph{degeneracy} of $C_k$. The expression for $\mu_m(C)$ is then obtained by summing over all $m_1$ from $N$ to $m$ and over all combinations of degeneracies such that their sum $\Vert\epsilon\Vert$ is equal to $m_2=m-m_1$,
\begin{align}
\mu_m(C)&=\sum_{m_1=N}^{m}(-1)^{m-m_1}\left(\prod_{i=1}^{m_1}\prod_{j=2}^{x(i)}\omega_{j-1}\right)\nonumber\\
&\sum_{\Vert\epsilon\Vert=m-m_1}\sideset{}{'}\sum_{t(x)}\prod_{k=1}^{l}\frac{1}{[e_0(C_k)]^{1+\epsilon_k}},
\label{mu_m}
\end{align}
where we remind that $l=x(1)+\dots+x(m_1)$, $\omega_L=\beta$ and $\Vert\epsilon\Vert=\epsilon_1+\dots+\epsilon_l$. The primed sum means that the value of $(L+1)$-th box in the $k$-th row of $t(x)$ must not be equal to $k(L+1)$ for all $k=1,\dots,m_4=m_1-N$ in order avoid visiting the empty lattice configuration. Together with Eqs.~(\ref{Pseries}), (\ref{cn_c0}) and (\ref{c0_recurrence}), this result concludes our power series in $\alpha$ for the inhomogeneous TASEP. 

\textit{Example}. Let us compute $\mu_2(1_2)$ for the system of size $L=3$, so that $m=2$ and $N=1$. Following Eq.~(\ref{mu_m}), the first option is to have two particles entering ($m_1=2$), no rule $2$ applied ($m_2=0$) and one particle leaving ($m_4=1$) so that $x(1)=L+1=4$ and $x(2)=2$ for which the corresponding Young diagram is
\begin{equation*}
\young(~~~~,:~~)
\end{equation*}
There are five fillings of this diagram with integers $1,\dots,l=x(1)+x(2)=6$ corresponding to sequences $111122$ ,$111212$, $111221$, $112112$ and $112121$, but the sequence $111122$ is excluded because it revisits the empty configuration. The second option is to have one particle entering ($m_1=1$), rule $2$ applied once ($m_2=1$) and no particles leaving ($m_4=0$), so that $x(1)=2$ for which the Young diagram is
\begin{equation*}
\yng(2)
\end{equation*}
with only one filling corresponding to the sequence $11$. Of the two configurations visited in the sequence $11$, $C_1=1_1$ and $C_2=1_2$, only $C_2$ has $\tau_1=0$, so that $\epsilon_1=0$ and $\epsilon_2=1$. The final expression for $\mu_2(1_2)$ has five contributions in total, the first four are from the first option and the last one is from the second option,
\begin{align}
&\mu_2(1_2)=\frac{\omega_{1}^{2}\omega_{2}\omega_3}{\omega_1\omega_2\omega_3(\omega_1+\omega_3)(\omega_1+\omega_4)(\omega_2+\omega_4)}\nonumber\\
&\quad+\frac{\omega_{1}^{2}\omega_{2}\omega_3}{\omega_1\omega_2\omega_3(\omega_1+\omega_3)(\omega_2+\omega_3)(\omega_2+\omega_4)}\nonumber\\
&\quad+\frac{\omega_{1}^{2}\omega_{2}\omega_3}{\omega_1\omega_{2}^{2}(\omega_1+\omega_3)(\omega_1+\omega_4)(\omega_2+\omega_4)}\nonumber\\
&\quad+\frac{\omega_{1}^{2}\omega_{2}\omega_3}{\omega_1\omega_{2}^{2}(\omega_1+\omega_3)\omega_3(\omega_2+\omega_4)}-\frac{\omega_1}{\omega_1\omega_{2}^{2}}.
\end{align}

\subsection{Power series in $\beta$}
\label{small_beta}

Rather than repeating the calculations from the previous Section, we can use the fact that the model is symmetric with respect to the following symmetry transformations
\begin{subequations}
	\begin{align}
	& 1_i \longleftrightarrow 0_i\\
	& \alpha\longleftrightarrow \beta\\
	& i\longleftrightarrow L-i+1.
	\end{align}
\end{subequations}
The first relation is the particle-hole symmetry which replaces particles with holes. Since holes move in the opposite direction, we also have to reverse the direction of the flow, which is done by the second and third relations. For example, applying these transformations to Eq.~(\ref{J_3}) gives the first three terms in the power series of $J$ in the exit rate $\beta$
\begin{align}
J     & =\beta-\frac{1}{\omega_L}\beta^2+\bigg(\frac{1}{\omega_L}-\frac{1}{\omega_{L-1}}\bigg)\Bigg[\frac{1}{\omega_{L-1}}+\sum_{j=1}^{L-2}\bigg(\frac{1}{\omega_j}\nonumber\\
&+\delta_{j,1}\frac{1}{\omega_1}\bigg)\prod_{q=j}^{L-2}\frac{\omega_q}{\omega_L+\omega_q}\Bigg]\beta^3+O(\beta^4).
\label{J_3_beta}
\end{align}
\subsection{Other cases}
\label{other_cases}

In Section \ref{small_alpha} we assumed that $\alpha$ is much smaller than any of the rates $\omega_1,\dots,\omega_L$ and $\beta$. The results of Section \ref{small_alpha} do not apply to the case in which one (or more) of these rates is equal to $\alpha$. Instead, each of these cases has to be studied separately starting from the zeroth order. For example, there are $L+1$ configurations for which $f_0(C)\neq 0$ in the power series in $\alpha=\beta$, which are $C=1_i\dots 1_L$ for $i=1,\dots,L$ and the empty lattice $C=\emptyset$. 

Another case that we do not cover in this work is the power series in one of the rates $\omega_i$. The simplest scenario is if $\omega_i\ll\alpha,\beta,\omega_j$ for all $j\neq i$ (the ``slow'' site problem). A special case in which $\omega_i\ll 1$ and $\omega_j=1$ for all $j\neq i$ has been studied in~\cite{SzavitsNossan13}. Other scenarios in which one or more rates $\alpha,\beta$ and $\omega_j$ for $j\neq i$ are equal to $\omega_i$ are in general more difficult to analyse and are beyond the scope of this paper.

\section{Conclusion}
\label{conclusion}

We have presented an analytic method for finding the steady state of the inhomogeneous TASEP as a power series in the entrance rate $\alpha$, assuming that $\alpha$ is small. This is the case of mRNA translation for which the rate of ribosome recruitment $\alpha$ is typically one or two orders of magnitude smaller than the hopping rates $\omega_i$~\cite{Ciandrini13,Duc18}. 

A practical advantage of our method is that the steady state probability $P(C)\sim O(\alpha^N)$, where $N$ is the number of particles in configuration $C$. Thus the computation of low-order terms is needed for only a small fraction of all configurations. In this paper we performed the expansion up to the second order, which allowed us to find an analytic expression for the density $\rho$ up to the quadratic order and for the particle current $J$ up to the cubic order in $\alpha$. We also presented an algorithm for computing higher-order terms recursively for small $n$.

The exact steady state of the inhomogeneous TASEP is a long outstanding problem in nonequilibrium statistical physics. Additionally, the mean-field approximation can only be found numerically~\cite{Shaw04}. Our analytic method reduces this gap and reveals how density and particle current depend on the particular sequence of hopping rates. The outstanding challenge for future work is to get practical information from the higher-order terms.

Our framework, which is applied here to the standard TASEP, can be extended to other models based on the exclusion process. In Ref.~\cite{SNCR18} we applied this method to a two-state~\cite{Klumpp08, Ciandrini10} TASEP with extended particles of size $\ell=10$~\cite{Shaw04}, which allowed us to predict mRNA sequence determinants of the rate of translation in yeast. In future, this approach may help to understand the role of codon usage bias, which remains one of the major unanswered questions in molecular biology, and could also be exploited in bioengineering.

\begin{acknowledgments}
	JSN was supported by the Leverhulme Early Career Fellowship under grant number ECF-2016-768. MCR was supported by the Biotechnology and Biological Sciences Research Council (BBSRC) BB/N017161/1 and the Scottish Universities Life Sciences Alliance. LC would like to thank the CNRS for having granted him a ``demi-d\'el\'egation'' (2017-18).
\end{acknowledgments}

\appendix
\section{Derivation of Eq.~(\ref{cn})}
\label{appendix_a}

Starting from Eq.~(\ref{Pseries}), we multiply the denominator of $P(C_i)$ by the power series on r.h.s. which gives
\begin{equation}
\sum_{n=0}^{K(C_i)}f_n(C_i)\kappa^n=\sum_{j=1}^{\mathcal{N}}\sum_{p=0}^{\infty}\sum_{m=0}^{K(C_j)}c_p f_m(C_j)\kappa^{m+p}
\end{equation}
Next, we introduce
\begin{equation}
K_0=\underset{j}{\textrm{max}}\{K(C_j)\}
\end{equation}
and set $f_m(C_j)=0$ for $K(C_j)<m\leq K_0$, so that
\begin{equation}
\sum_{n=0}^{K(C_i)}f_n(C_i)\kappa^n=\sum_{p=0}^{\infty}\sum_{m=0}^{K_0}c_p b_m\kappa^{m+p},
\end{equation}
where $b_m$ is defined in Eq.~(\ref{bn}). We now define $n=m+p$ so that
\begin{equation}
\sum_{n=0}^{K(C_i)}f_n(C_i)\kappa^n=\sum_{n=0}^{\infty}\left(\sum_{m=0}^{\textrm{min}\{n,K_0\}}c_{n-m} b_m\right)\kappa^{n}.
\end{equation}
Since by definition $K(C_i)\leq K_0$, for $0\leq n\leq K(C_i)$ we write
\begin{eqnarray}
f_n(C_i)&=&\sum_{m=0}^{n}b_m c_{n-m}=\sum_{m=0}^{n}b_{n-m} c_{m}\nonumber\\
&=&b_0 c_n+\sum_{m=0}^{n-1}b_{n-m} c_{m},
\end{eqnarray}
which after rearranging gives Eq.~(\ref{cn}). Similarly, we write
\begin{eqnarray}
0&=&\sum_{m=0}^{\textrm{min}\{n,K_0\}}b_m c_{n-m}+b_0 c_n\nonumber\\
&+&\sum_{m=1}^{\textrm{min}\{n,K_0\}}b_m c_{n-m},
\end{eqnarray}
for $n>K(C_i)$, which after rearranging gives Eq.~(\ref{cn_gt_K}).

\section{An algorithm for finding $c_n(C)$ recursively for small $n$}
\label{appendix_b}

Our starting position for computing $c_n(C)$ is Eq.~(\ref{cn}) for $n\leq K(C)$, which reads
\begin{equation}
c_n(C)=\frac{f_n(C)-\sum_{m=0}^{n-1}b_{n-m}c_{m}(C)}{b_0}.
\label{cn_appendix}
\end{equation}
In Eq.~(\ref{fn_f0}) we show that $f_n(C)$ can be written as a linear combination of unknown empty-lattice coefficients $f_{0}(\emptyset),\dots,f_{N(C)}(\emptyset)$,
\begin{equation}
f_n(C)=\sum_{m=N(C)}^{n}\mu_{m}(C)f_{n-m}(\emptyset), 
\label{fn_f0_appendix}
\end{equation}
where $N(C)$ is the number of particles in configuration $C$. Most importantly, we also show that $c_n(C)$ does not depend on any of these empty-lattice coefficients, which means that the numerator in Eq.~(\ref{cn_appendix}) must be proportional to $b_0=f_0(\emptyset)$. In other words, none of the contributions to $f_n(C)$ and $b_{n-m}$ from any of $f_{m}(\emptyset)$ other than $f_0(\emptyset)$ matter; they all cancel each other out. For all practical purposes, that is numerically equivalent of setting the value of all $f_{n-m}(\emptyset)$ to zero except for $f_0(\emptyset)$, whose value is set to $1$.

The algorithm for computing $c_n(C)$ first finds all $f_n(C)$ and then $c_n(C)$ according to Eq.~(\ref{cn_appendix}), assuming that we know $f_{n-1}(C)$ for all $C$ and $b_{n-m}$ for all $m=0,\dots,n-1$. The first step is to compute $f_n(C)$ for configurations that have precisely $n$ particles (we remind that $f_n(C)=0$ if $N(C)>n$), starting from the configuration in which $n$ particles occupy the first $n$ sites, $C=1_{1} 1_{2} \dots 1_{n}$. In that case Eq.~(\ref{recurrence}) reads
\begin{equation}
f_n(1_1\dots 1_n)=\frac{1}{\omega_{n}}f_{n-1}(1_2\dots 1_n).
\end{equation}
The next step is to compute $f_n(1_1 1_{x(2)}\dots 1_{x(n)})$ for all $2\leq x(2)<\dots<x(n)\leq L$, using the following recurrence equation,
\begin{align}
& f_n(1_1 1_{x(2)}\dots 1_{x(n)})=\frac{1}{e_0}f_{n-1}(1_{x(2)}\dots 1_{x(n)})\nonumber\\
& \quad + \sum_{k=2}^{n}\frac{\omega_{x(k)-1}}{e_0}f_{n}(1_1\dots 1_{x(k)-1}\dots 1_{x(n)}),
\label{recurrence_appendix}
\end{align}
where $e_0(x)$, $x=(x(1),\dots,x(n))$ is given by
\begin{equation}
e_0(x)=\sum_{k=1}^{n-1}(1-\delta_{x(k),x(k+1)-1})\omega_{x(k)}+\omega_{x(n)}
\end{equation}
In Eq.~(\ref{recurrence_appendix}) we have set the value of $f_n(1_{x(1)}\dots 1_{x(n)})$ to zero if any two neighbouring particle coordinates $x(k),x(k+1)$ become equal, 
\begin{equation}
f_n(\dots 1_{x(k)} 1_{x(k)}\dots)=0,
\end{equation}
which ensure that the exclusion principle is satisfied. We note that since $f_n(C)=0$ if the number of particles in $C$ is larger than $n$, there is no term in Eq.~(\ref{recurrence_appendix}) that introduces a new particle from the right reservoir, i.e. $f_n(1_1\dots 1_{x(n)}1_{x(n+1)})=0$. That is the reason why we can solve Eq.~(\ref{recurrence_appendix}) recursively.

After we have computed all $f_n(1_1 1_{x(2)}\dots 1_{x(n)})$, we find all $f_n(1_2 1_{x(2)}\dots 1_{x(n)})$ using the following recurrence relation
\begin{align}
f_n(1_2 1_{x(2)}\dots 1_{x(n)}) &= \sum_{k=1}^{n}\frac{\omega_{x(k)-1}}{e_0}\nonumber\\
& \times f_{n}(\dots 1_{x(k)-1}\dots).
\label{recurrence2_appendix}
\end{align}
We repeat this procedure for all $x(1)=2,\dots,L-n$, until we have computed all $n$-particle coefficients $f_n(1_{x(1)}\dots 1_{x(n)})$.

In the next step we consider configurations with $n-1$ particles, starting from the configurations in which particles occupy the first $n-1$ sites. The only difference compared to the previous step is that now we can also move a particle from the right reservoir,
\begin{align}
& f_n(1_1 1_{x(2)}\dots 1_{x(n-1)})=\frac{1}{e_0}f_{n-1}(1_{x(2)}\dots 1_{x(n-1)})\nonumber\\
& \quad+\sum_{k=2}^{n-1}\frac{\omega_{x(k)-1}}{e_0}f_{n}(1_1\dots 1_{x(k)-1}\dots 1_{x(n-1)})\nonumber\\
& \quad+\frac{\omega_{L}}{e_0}f_{n}(1_1\dots 1_{x(n-1)1_{L}}).
\label{recurrence3_appendix}
\end{align}
The last term contains $n$ particles and has been computed in the previous step. A similar recurrence relation can be written for all other terms with $n-1$ particles. These steps are then repeated for $n-2$ particles, $n-3$ particles and so on, until we reach configurations with only one particle. In particular, the equation for $f_n(C=1_1)$ is given by
\begin{equation}
f_n(1_1)=\frac{1}{\omega_1}f_{n-1}(\emptyset)+\frac{\omega_L}{e_0}f_{n}(1_1 1_{L}),
\end{equation}
in which we set the value of $f_{n-1}(\emptyset)$ to zero, as we discussed before.

Finally, after we have found all non-zero $f_n(C)$, we compute $b_n$ from Eq.~(\ref{bn}) and finally $c_n(C)$ from Eq.~(\ref{cn_appendix}). 


\end{document}